\documentclass[11pt,aps,longbibliography]{revtex4-1}
\usepackage{amsmath,graphicx,subfigure,bm,xcolor,mathtools,newtxtext,soul}
\usepackage{braket}
\usepackage{ulem}
\usepackage{cancel}   
\usepackage[ bookmarks=true, colorlinks, linkcolor=blue, urlcolor=blue, citecolor=blue, plainpages=false, pdfpagelabels, final, breaklinks=true ]{hyperref}
\usepackage{ulem}

\begin{document}

\title{Quantum States Protection under Environmental Noise}

\author{Kai Wang$^{1}$ and Zhen-Yang Peng$^{2}$}
\address{$^{1}$ Interdisciplinary Artificial Intelligence Research Institute, Wuhan College, Wuhan 430212,China}
\address{$^{2}$ Wilczek Quantum Center, School of Physics and Astronomy, Shanghai Jiao Tong University, Shanghai 200240, China}
\email{pengzhenyang@sjtu.edu.cn}

\begin{abstract}
All realistic quantum systems are inevitably in contact with the environment. Suppressing the impact of environmental noise is a critical challenge in cutting-edge quantum technologies. 
In this work, we introduce and systematically analyze a scheme for the protection of quantum states against amplitude-damping (AD) noise based on the circuit structure called the quantum filter. 
Filtration circuits employing single- and multi-control qubits are examined, and their capability to enhance state protection fidelity while preserving a high success probability is discussed. 
Moreover, for many‑body qubit states, those with a fixed quantum Hamming weight can be perfectly protected against AD noise, whereas states with the largest Hamming weight difference set a lower bound on the achievable protection fidelity. 
Our work provides a resource-efficient route
for quantum state protection without requiring full quantum error correction.
\end{abstract}


\maketitle

\section{Introduction}

Quantum information processing is a rapidly developing research area that focuses on the generation, manipulation, and readout of quantum states~\cite{Kimble_quantuminternet,quantumchannel_rmp_2014,QEC_rmp_2017,Sciarrino_review_2019,Blais_natphys_2020}. However, any quantum states rely on physical platforms such as trapped ions, superconducting circuits, neutral atoms, and other hybrid quantum devices are in contact with the environment~\cite{Wineland_science_2009,Wendin_review_2017,Quantumsensing_rmp_2017,Monroe_rmp_2021,Wallraff_RMP_2021_cQED}. 
Environment-induced decoherence in quantum systems destroys quantum effects such as coherence and entanglement, posing a fundamental limitation to the further exploitation of quantum states. Therefore, mitigating the impact of environmental noise is crucial for state-of-the-art quantum technologies and beyond~\cite{Preskill_NISQ_2018,Nash_QST_2020,NISQ_rmp_2022}.

In qubit systems, the dominant environmental noise that induces energy relaxation ($T_1$) is commonly modeled by the amplitude damping (AD) channel~\cite{book_nielsen_chuang_quantum_2010,ad_code_1997,ad_code_2013}. 
Unlike depolarizing or Pauli channels, the AD channel captures the ubiquitous energy relaxation processes, such as spontaneous emission, which cause an irreversible transfer of population from the excited state $\ket{1}$ to the ground state $\ket{0}$. 
This dissipative dynamics imparts an irreversible, non-unitary character that cannot be fully reproduced by simple Pauli noise models obtained via Pauli twirling~\cite{Cai_npj_2020}, and consequently prevents recovery of the initial quantum state through simple operations.

Several strategies to protect the quantum states against AD noise have been explored both theoretically and experimentally in recent years. 
For instance, dynamical decoupling has been applied to preserve quantum states~\cite{Cleland_prl_2022,Retzker_prx_2023} and to protect quantum gates~\cite{Gong_pra_2025}.
Weak measurement and its reversal have been demonstrated to enable high-fidelity entanglement swapping~\cite{Harraz_2025} and state protection~\cite{Yacoby_prl_2010,PanJW_pra_2017}.
Auxiliary systems with flexible control provide another means for the effective cancellation of decoherence~\cite{Bernien_science_2023,Benatti_prl_2026}. 
These strategies generally require additional controls, interactions, or measurements on the target quantum systems to achieve state protection.
By contrast, a distinct class of strategies exploits the inherent symmetries of the system–environment interaction,
e.g., decoherence-free subspaces can be exploited for noise-resilient quantum control and information protection~\cite{Yuasa_pra_2016,Holland_pra_2022}. 
Symmetry verification has likewise been proposed as a quantum error mitigation technique aimed at reducing the effects of noise~\cite{symmetryverification_NC_2020,symmetryverification_quantum_2021,symmetryverification_prxquantum_2025}. 
Furthermore, once suitable reference states are chosen, the Petz recovery map can provide effective protection of the initial quantum states~\cite{Petzmap_physletta_2024,Petzmap_prl_2025,Petzmap_pra_2025}.

In this paper, we have discussed a strategy for protecting quantum states against AD noise via the circuit structure called the quantum filter. Starting from single-qubit states, we demonstrate that the filtration structure enhances fidelity under AD noise, outperforming the control-swapped structure that is widely used in error mitigation. 
The results of Bell-state protection show that the state $\ket{\Phi^{\pm}}$ under AD noise can be exactly recovered after the filtration circuit, providing perfect protection. Based on that, we then discussed the protection for many-body qubit states via filtration circuits employing single- and multi-control qubits. We have shown that states with fixed quantum Hamming weights, such as general $W$ states~\cite{Wstate_pra_2008} and Dicke states~\cite{symmetricstate_review_2026}, can be perfectly protected against AD noise, whereas GHZ states, which possess the maximal quantum Hamming distance, correspond to the lower bound of protection under filtration circuits. Meanwhile, the protection of general superposition states is numerically analyzed.

\section{Quantum filter for AD noise}

The quantum filter is a class of circuits designed for error suppression~\cite{error_filter_prl_2023,purify_filter_channel2024}. Its typical structure comprises the target quantum channel, control operations, and measurements on the control qubits. The implementation of a quantum channel inevitably introduces noise, which leads to errors or decoherence in the quantum state. The control operations serve to decompose the channel output into two distinct components. Thus, by post-selecting on the measurement outcome of the control qubit, a conditional output state with less error is obtained, thereby achieving error suppression.

The single-qubit AD channel is given by
\begin{equation}
    \mathcal{E}^{\text{AD}}(\cdot)=E_0 \cdot E_0^{\dagger}+E_1 \cdot E_1^{\dagger}
\end{equation}
with Kraus operators
\begin{equation}
    \begin{split}
    E_0 &= \ket{0}\bra{0} + \sqrt{1-\gamma}\ket{1}\bra{1},\\
    E_1 &= \sqrt{\gamma}\ket{0}\bra{1},
    \end{split}
\end{equation}
where $\gamma$ is the damping rate. 
Here,  $E_0$ describes the no‑jump evolution, corresponding to the attenuation of the excited state, 
whereas $E_1$ represents the quantum jump from the excited state to the ground state.

\begin{figure}[h]
    \centering
    \includegraphics[width=0.9\columnwidth]{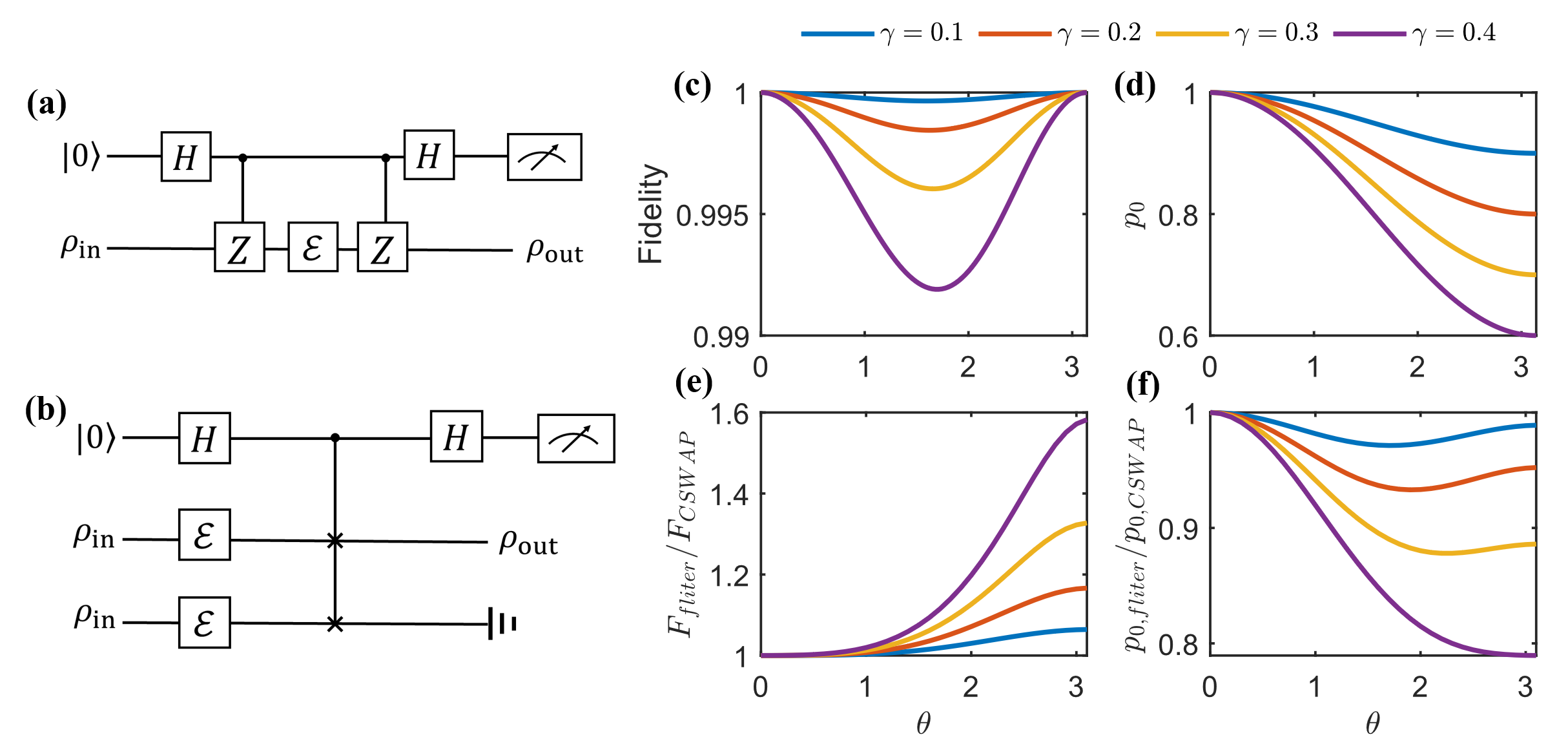}
    \caption{
    (a) Sketch of the circuit for quantum state protection via a quantum filter under AD noise. (b) Circuit for the controlled-SWAP operation. In both (a) and (b), 
    $\mathcal{E}$ represents the AD noise channel, and the first qubit serves as the control qubit. 
    (c) and (d) show the fidelity and the success probability for single-qubit quantum filtration, respectively. 
    (e) Ratio of the fidelities between quantum filtration and the CSWAP operation. (f) Ratio of the success probabilities between quantum filtration and the CSWAP operation.}
    \label{fig:fig1}
\end{figure}

Due to the commutation and anti‑commutation relations of the above Kraus operators with the Pauli Z operator, namely
\begin{equation}
    [E_0,Z]=0, \qquad \{ E_1,Z \}=0,\label{eq:commutation}
\end{equation}
the control operation for the quantum filter that suppresses AD noise can be implemented using controlled‑Z (CZ) gates.
The corresponding circuit structure is shown in~\ref{fig:fig1}(a).
As a result, the total output state before measurement has the form
\begin{equation}
    \begin{split}
    \rho_{\text{tot}}&=\frac{1}{2} \left\{ \ket{+}_c \bra{+}\otimes\mathcal{E}^{\text{AD}}(\rho_{\text{in}}) + \ket{+}_c \bra{-}\otimes \mathcal{E}^{\text{AD}}(\rho_{\text{in}}Z)Z \right. \\
      &\left.+ \ket{-}_c \bra{+}\otimes Z\mathcal{E}^{\text{AD}}(Z\rho_{\text{in}}) + \ket{-}_c \bra{-}\otimes Z\mathcal{E}^{\text{AD}}(Z\rho_{\text{in}}Z)Z \right\} ,
    \end{split}
\end{equation}
where $\ket{\pm}_c=(\ket{0}_c \pm \ket{1}_c)/\sqrt{2}$ are the eigenstates of the Pauli X operator for the control qubit.
The final output state is obtained by post‑selection on the measurement outcome of the control qubit. 
The probability of obtaining a specific measurement outcome is given by
\begin{equation}
    p_{\text{meas}} = \text{Tr}
    \left\{\bra{m}_c
    \rho_{\text{tot}}\ket{m}_c\right\},
\end{equation}
where $\ket{m}_c$ denotes the measurement basis state of the control qubit.
Conditioned on this outcome, the output state after post‑selection takes the form
\begin{equation}
    \rho_{\text{out}}=\bra{m}_c \rho_{\text{tot}}\ket{m}_c/p_{\text{meas}}. 
\end{equation}
Since the $E_1$ Kraus operator in the AD channel describes the damping process from $\ket{1}$ to $\ket{0}$ with rate $\gamma$, 
the filter structure has been designed to suppress the influence of damping that is induced by $E_1$ by exploiting its anti-commutation relation given in Eq.~\eqref{eq:commutation}.
The fidelity between the final output state $\rho_{\text{out}}$ and the input state $\rho_{\text{in}}$ defined as 
\begin{equation}
    F(\rho_{\text{out}},\rho_{\text{in}}) = \text{Tr}\sqrt{\sqrt{\rho_{\text{out}}}\rho_{\text{in}} \sqrt{\rho_{\text{out}}}}
\end{equation}
can be used to quantify the protection efficiency of the initial state. 
If the initial state is a pure state $\ket{\psi}$, 
then the fidelity can be conveniently simplified to $F=\sqrt{\bra{\psi}\rho_{\text{out}}\ket{\psi}}$.

\subsection{Single qubit}
We first discuss the simplest case of the filtration effect, 
in which the circuit is composed of only one target qubit and one control qubit.
The circuit filters out the damping component of the AD noise 
once the measurement outcome $\ket{0}_c$ is obtained, 
and the final output state contains only the $E_0$ term:
$\rho_{\text{out}}=E_0 \rho_{\text{in}} E_0^{\dagger} /p_0$. 
If the initial state is $\ket{\psi} = \cos{\frac{\theta}{2}}\ket{0} + e^{\text{i}\phi}\sin{\frac{\theta}{2}}\ket{1}$, 
the final output state conditioned on the measurement outcome 
$\ket{0}$ is
\begin{equation}
    \rho_{\text{out}} =\frac{2}{2{-}\gamma {+}\gamma \cos{\theta}} 
    \begin{pmatrix}
    \cos^2{\frac{\theta}{2}}  & \frac{1}{2}e^{-\text{i}\phi}\sqrt{1{-}\gamma} \sin{\theta} \\
    \frac{1}{2}e^{\text{i}\phi}\sqrt{1{-}\gamma} \sin{\theta} & (1{-}\gamma)\sin^2{\frac{\theta}{2}}\end{pmatrix},
\end{equation}
with the probability of obtaining $\ket{0}$ on the control qubit
\begin{equation}
    p_0 {=} (2{-}\gamma {+}\gamma \cos{\theta})/2,
\end{equation}
and the corresponding fidelity is then
\begin{equation}
  F =\sqrt{\frac{2\,\cos^4{\left(\frac{\theta }{2}\right)} +2(1-\gamma )\,{\sin^4 \left(\frac{\theta }{2}\right)} + \sin^2 {\theta}\,\sqrt{1-\gamma }}{\gamma \,\cos\theta -\gamma +2}}.
\label{eq:fidelity_1qubit}
\end{equation}
While the controlled‑SWAP (CSWAP) gate, 
as illustrated in Fig.~\ref{fig:fig1}(b), 
is widely used for quantum error mitigation~\cite{Jorge_prl_2023,Jorge_pra_2023}
and quantum state and channel purification~\cite{purify_state_2022,purifychannel2022,purifychannel2025},
we demonstrate that 
our AD‑noise filtering scheme provides a resource-efficient and higher-fidelity advantage in protecting quantum information under AD noise.

The numerical results for the fidelity and success probability of single-qubit quantum filtration are shown in Fig.~\ref{fig:fig1}(c) and (d),
while their direct comparison with the CSWAP operation is provided in Fig.~\ref{fig:fig1}(e) and (f) via the corresponding ratios.
It can be noticed that for the filtering operation, 
the initial state $\ket{1}$ can be perfectly protected with unit fidelity $F=1$, even in the presence of AD noise. 

In sharp contrast, the performance of CSWAP operations is fundamentally constrained by the intrinsic dissipative dynamics of AD noise, as the continuous energy exchange with the environment drives the system toward the ground state, depletes the excited‑state population, and thereby sets an inherent upper bound on the achievable fidelity enhancement.
Moreover, despite being designed for efficient error suppression~\cite{streaming_purification_2025_Quantum}, the streaming CSWAP operation suffers from an exponential growth in resource overhead, i.e., the number of required copies of the initial state explodes as the recursion depth increases.
Therefore, our filtering scheme is free from such a resource explosion, offering a resource‑efficient and fidelity-superior alternative for protecting quantum information under AD noise.

\subsection{Bell State}
The four Bell states are maximally entangled two-qubit states with important applications in quantum technologies. 
Following the discussion of single-qubit filtration, we turn to the quantum filter for Bell-state protection.
Without loss of generality, we assume that each qubit is subject to independent AD noise with the same damping rate.
The overall Kraus operators are then given by the tensor products of the local single-qubit operators $E_0$ and $E_1$.
The filtration retains only symmetric Kraus operators while filtering out all antisymmetric ones when the measurement outcome $\ket{0}_c$ is obtained, resulting in $\rho_{\text{out}}\propto E_{00}\rho_{\text{in}}E^{\dagger}_{00}+E_{11}\rho_{\text{in}}E^{\dagger}_{11}$, with the overall two-qubit Kraus operators are $E_{ij}=E_{i}\otimes E_{j}$.
Since each $E_1$ provides an anti-symmetric contribution that will be filtered out,
the filtration removes any term containing an odd number of $E_1$ factors; this explains the perfect protection of $\ket{1}$ in the single‑qubit case. In contrast, the two‑mode damping operator $E_{11}$, being the product of two antisymmetric components, is symmetric and therefore survives the filtering process, contributing to the final state. For two‑qubit filtration, a phenomenological prediction is that this additional $E_{11}$ term contains the total damping effect, thereby reducing the fidelity.

Taking the Bell states $\ket{\Psi^{\pm}}=(\ket{00}\pm \ket{11})/\sqrt{2}$ as the input, one can calculate the post-selected output density matrix as
\begin{equation}
    \rho_{\text{out},\ket{\Psi^{\pm}}} = \frac{1}{2{-}2\gamma{+}2\gamma^2}\begin{pmatrix}
        1{+}\gamma^2 & 0 & 0 & \pm (1{-}\gamma) \\
        0 & 0 & 0 & 0 \\
        0 & 0 & 0& 0 \\
        \pm(1{-}\gamma) & 0 & 0 & (1{-}\gamma)^2
    \end{pmatrix},
\end{equation}
with the corresponding fidelity
\begin{equation}
    F_{\ket{\Psi^{\pm}}} = \sqrt{ \frac{2-2\gamma+\gamma^2}{2-2\gamma+2\gamma^2} }.\label{eq:Bell_Psi}
\end{equation}
We now turn to the $\ket{\Phi^{\pm}}$ Bell states. Remarkably, when the input state is $\ket{\Phi^{\pm}}$, post-selection yields an output state that is strictly identical to the input. This indicates that for the $\ket{\Phi^{\pm}}$ Bell states, our filtration circuit provides perfect protection against AD noise. In other words, our scheme fully corrects the AD errors for these states, preserving them with unit fidelity upon successful post-selection. We will discuss this perfect protection against AD noise in depth for multi-qubit quantum states in the following section.

\section{Multi-qubit quantum states}
We have demonstrated that the Bell states $\ket{\Phi^{\pm}}$ can be perfectly protected by our filtration under AD noise, 
whereas the states $\ket{\Psi^{\pm}}$ remain susceptible to damping. 
We note that the essential distinction between $\ket{\Phi^{\pm}}$ and $\ket{\Psi^{\pm}}$ can be characterized by the quantum Hamming weight~\cite{book_quantumalgorithms_2025,Wilde_pra_2024}:
$\ket{\Phi^{\pm}}$ possess a fixed quantum Hamming weight of $1$, while $\ket{\Psi^{\pm}}$ are superpositions of states with different Hamming weights. 
Inspired by the protection of the Bell states,  
we now extend our investigation to many-body systems. 
In this context, 
quantum states categorized by their Hamming weight provide a natural framework for studying the protection of multi-qubit states under our quantum filtration scheme.

For $N$-qubit states, the total AD noise can be denoted as
\begin{equation}
    \mathcal{E}^{\text{AD}}(\cdot) = \sum_{\alpha=1}^{2^N}K_{\alpha}\cdot K_{\alpha}^{\dagger}. \label{eq:N_ADchannel}
\end{equation}
Here $K_{\alpha}$ indicates the total Kraus operator defined as the tensor product $ E_{\alpha_1} \otimes E_{\alpha_2} \cdots \otimes E_{\alpha_N} $,  
where each $E_{\alpha_i} \in \{E_0,E_1\}$ is selected according to the $i$-th binary digit of the index $\alpha$~(i.e., $\alpha_i \in \{0,1\}$ determines whether $E_0$ or $E_1$ is used). For simplicity, we define $P(N, m)$ as the set of all distinct permutations of a binary sequence consisting of $m$ ones followed by $N-m$ zeros. In other words, $P(N, m)$ contains all binary strings of length $N$ with exactly $m$ ones. Therefore, the Eq.~\eqref{eq:N_ADchannel} can be rewritten as
\begin{equation}
    \mathcal{E}^{\text{AD}}(\cdot) = \sum_{m=0}^N \sum_{k=1}^{\binom{N}{m}}
   E_{P_{k} {(N,m)}}\cdot E^{\dagger}_{P_{k} {(N,m)}},
\end{equation}
Here, for each $m$, the index $k$ runs over the $\binom{N}{m}$ elements of the set $P {(N,m)}$.
An example of $E_{P_{k} {(N,m)}}$ for $N=4$ is given in Table \ref{tab:my_label}.
\begin{table}[htbp]
    \centering
    \begin{tabular}{c|c}
    \hline
        $E_{P(N,0)}$ & $E_{0000}$  \\
        \hline
        $E_{P(N,1)}$ & $E_{0001}, E_{0010}, E_{0100}, E_{1000}$ \\
        \hline
        $E_{P(N,2)}$ & $E_{0011}, E_{0101}, E_{1001}, E_{0110}, E_{1010}, E_{1100}$ \\
        \hline
        $E_{P(N,3)}$ & $E_{0111}, E_{1011}, E_{1101}, E_{1110}$ \\
        \hline
        $E_{P(N,4)}$ & $E_{1111}$ \\
        \hline
    \end{tabular}
    \caption{An example for the notations of $N=4$ Kraus operators.}
    \label{tab:my_label}
\end{table}
The post-selection on the measurement outcome $\ket{0}$ of the control qubit leads to the corresponding output states
\begin{equation}
    \rho_{\text{out}} = \frac{1}{4p_0} \left( \mathcal{E}[\rho_{\text{in}}] + \tilde{Z}\mathcal{E}[\tilde{Z}\rho_{\text{in}}] + \mathcal{E}[\rho_{\text{in}}\tilde{Z}]\tilde{Z} + \tilde{Z}\mathcal{E}[\tilde{Z}\rho_{\text{in}}\tilde{Z}]\tilde{Z} \right). \label{eq:state_out}
\end{equation}
Here $p_0$ is the success probability of obtaining the measurement result $\ket{0}$, $\tilde{Z}=\bigotimes_{i=1}^N Z_i$, and $Z_i$ is the Pauli-$Z$ operator acting on the $i$-th qubit. 
Based on the commutation/anti-commutation relation given in Eq.~\eqref{eq:commutation},
the filtration process for $N$-qubit states eliminates the Kraus operators 
that contain an odd number of $E_1$ factors, 
and the final post-selected output given in Eq.~\eqref{eq:state_out} can be further simplified as 
\begin{equation}
    \rho_{\text{out}} = \frac{1}{p_0} \sum_{m=0,2,4\dots}^{m\leq N} \sum_{k=1}^{\binom{N}{m}} E_{P_{k}(N,m)}\rho_{\text{in}} E^{\dagger}_{P_{k}(N,m)}.
    \label{eq:state_out_simplify}
\end{equation}

\subsection{States with fixed quantum Hamming weight}
We first consider a class of states $\ket{\psi} \in \mathcal{H}_{s,N}$, where $\mathcal{H}_{s,N}$ denotes the subspace spanned by the $N$-qubit states with Hamming weight $s$. 
Let $\{ \ket{\phi^{(i)}_{s,N}} \}$ be an orthonormal basis of $\mathcal{H}_{s,N}$, then for any fixed initial pure state $\ket{\psi_0} \in \mathcal{H}_{s,N}$,
any pure state $\ket{\psi} \in \mathcal{H}_{s,N}$ can be generated as
$\ket{\psi}=U_{s,N}\ket{\psi_0}$
via the unitary transformation $U_{s,N}=\exp{(-\text{i}\sum_{j,k} \theta_{j,k}\ket{\phi^{(j)}_{s,N}}\bra{\phi^{(k)}_{s,N}})}$, where $\theta_{k,j}^*=\theta_{j,k}$. 
Before further discussion, 
we point out that all input states
$\rho_{\text{in}} \in  \mathcal{D}(\mathcal{H}_{s,N}) $,
where \( \mathcal{D}(\mathcal{H}_{s,N}) \) denotes the set of density operators on the Hilbert space \( \mathcal{H}_{s,N} \),
share the same fidelity when subjected to the AD noise channel, 
Importantly, they also achieve the same fidelity 
after our filtration circuit (Fig.\ref{fig:fig1}a) conditioned on successfully measuring
$\ket{0}$
in the presence of the AD channel.
More precisely,
\begin{equation}
    \begin{split}
    F(\mathcal{E}^{\text{AD}}(\rho_{\text{in}}),\rho_{\text{in}})
    &=F(\mathcal{E}^{\text{AD}}(\ket{\phi^{(i)}_{s,N}}\bra{\phi^{(i)}_{s,N}}),\ket{\phi^{(i)}_{s,N}}\bra{\phi^{(i)}_{s,N}})
    \\
    F(S[\mathcal{E}^{\text{AD}}(\rho_{\text{in}})],\rho_{\text{in}})
    &=F(S[\mathcal{E}^{\text{AD}}(\ket{\phi^{(i)}_{s,N}}\bra{\phi^{(i)}_{s,N}})],\ket{\phi^{(i)}_{s,N}}\bra{\phi^{(i)}_{s,N}})
  \label{eq:fid_invariant}
    \end{split}
\end{equation}
where $S$ denotes the superchannel description of the filter. The detailed derivation is provided in Appendix~\ref{app:1}. 
This property suggests that the filtration circuit in Fig.~{\ref{fig:fig1}}(a) for any input state $\rho_{\text{in}} \in  \mathcal{D}(\mathcal{H}_{s,N}) $ 
can be equivalently analyzed by using any one of the eigenstates $\ket{\phi^{i}_{s,N}}$. 
Therefore, the fidelity between the input state with quantum Hamming weight $s$
and the post-selected output state as in Eq.~\eqref{eq:state_out_simplify} can be readily computed and takes the form
\begin{equation}
    F_{s,N} = \sqrt{\frac{(1-\gamma)^s}{p_0}},\label{eq:fid_fixHamming_1ctr}
\end{equation}
with the success probability of obtaining the measurement outcome $\ket{0}$
given by
\begin{equation}
    p_0 = \sum_{m=0,2,\dots}^{m\leq s} \binom{s}{m}(1-\gamma)^{s-m}\gamma^m.
    \label{eq:pro_suc}
\end{equation}
Eq.~\eqref{eq:fid_fixHamming_1ctr} reveals that the fidelity is independent of the total number of qubits
and depends primarily on the Hamming weight $s$.
More importantly, 
from the success probability expression in Eq.~\eqref{eq:pro_suc},
it can be calculated that all states with quantum Hamming weight $s=1$, 
e.g., Bell state $\ket{\Phi^{\pm}}$, and generalized W-states, 
can be perfectly protected, 
achieving a fidelity of $F=1$ with a success probability of $1-\gamma$.

\subsection{Superpositions of states with different Hamming weights}
According to the fidelity results presented for a single qubit in Eq.~\eqref{eq:fidelity_1qubit} and for the Bell states $\ket{\Psi^{\pm}}$ in Eq.~\eqref{eq:Bell_Psi}, it can be generalized that superposition states that involving components from different subspaces $\mathcal{H}_{s,N}$ cannot be perfectly protected by the filtration structure introduced above. 
Nonetheless, this structure can still provide efficient protection for such superposition states against the AD noise.
As an example, the GHZ states, for which the two coherent subspaces possess the maximum Hamming distance, are discussed analytically here, while the rest of the superposition cases are numerically calculated.
Consider the $N$-qubit GHZ state
\begin{equation}
    \ket{\text{GHZ}}_N = \frac{\ket{0}^{\otimes N} +\ket{1}^{\otimes N}}{\sqrt{2}}
\end{equation}
as the input state of the filtration circuit. 
The output state after post-selection can be calculated via Eq.~\eqref{eq:state_out_simplify}, which has the form
\begin{equation}
\begin{split}
    \rho_{\text{out,GHZ}} &= \frac{1}{2p_{0,\text{GHZ}}} \left\{ \ket{0}^{\otimes N}\bra{0}^{\otimes N} + (1-\gamma)^{N/2}(\ket{0}^{\otimes N}\bra{1}^{\otimes N}+\ket{1}^{\otimes N}\bra{0}^{\otimes N})\right. \\
    &\left. + \sum_{m=0,2,\dots}^{m\leq N}\sum_{\mathcal{P}} \gamma^{m} (1-\gamma)^{N-m}\mathcal{P} (\ket{0}^{\otimes m} \otimes \ket{1}^{\otimes (N-m)}) (\bra{0}^{\otimes m} \otimes \bra{1}^{\otimes (N-m)} )\mathcal{P} \right\},
    \label{eq:output_GHZ_1ctr}
\end{split}
\end{equation}
where $\mathcal{P}$ runs over all distinct permutations that yield different configurations of the $\ket{1}$ states across the $N$ qubits.
Here $p_{0,\text{GHZ}}$ denotes the success probability 
\begin{equation}
    p_{0,\text{GHZ}} = \frac{1}{2} + \frac{1}{2}\sum_{m=0,2\dots}^{m\leq N} \binom{N}{m}\gamma^m (1-\gamma)^{N-m}.
\end{equation}
The fidelity between the state in Eq.~\eqref{eq:output_GHZ_1ctr} and $\ket{\text{GHZ}}_N$ is then given by
\begin{equation}
    F_{\text{GHZ}} = \begin{cases}
    \frac{1}{2\sqrt{p_{0,\text{GHZ}}}} \sqrt{ \left[1+(1-\gamma)^{N/2}\right]^2 +\gamma^N }, & N~ \text{is even}\\
    \frac{1}{2\sqrt{p_{0,\text{GHZ}}}}\left[1+(1-\gamma)^{N/2}\right], & N~\text{is odd}.
    \end{cases}\label{eq:fidelity_GHZ_1ctr}
\end{equation}

\begin{figure}[h]
    \centering
    \includegraphics[width=0.8\columnwidth]{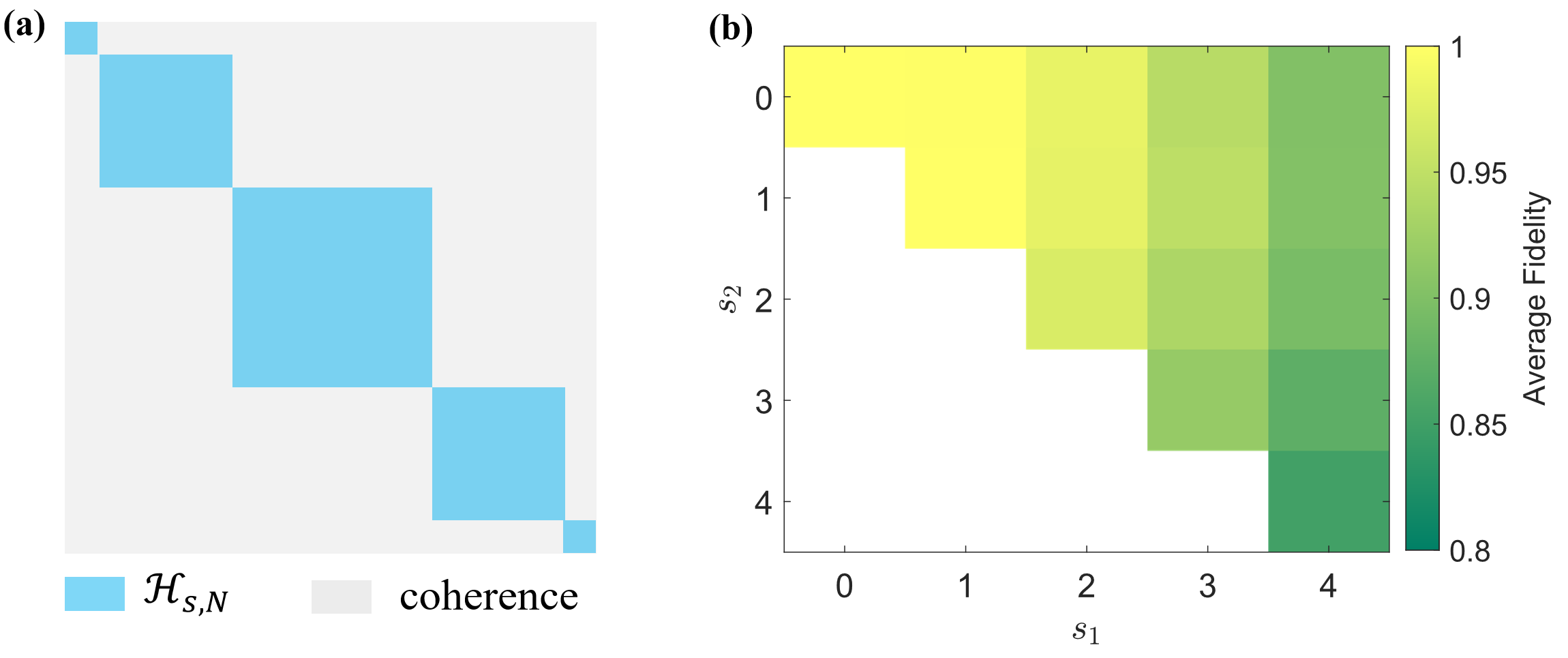}
    \caption{(a) Sketch illustrating the decomposition of the Hilbert space for $N=4$ qubits. The block-diagonal regions correspond to subspaces with a fixed quantum Hamming weight, while the off-diagonal part represents the subspace of superposition states that possess coherence between different quantum Hamming weights.
    (b) Average fidelity of bipartite superposition states with quantum Hamming weights $s_1$ and $s_2$. 
    The diagonal terms correspond to the cases described by Eq. (18).
    All average fidelities shown in this figure are evaluated at $\gamma=0.2$.}
    \label{fig:1ctr}
\end{figure}

The numerical results are shown in Fig.~\ref{fig:1ctr}. We have calculated the average fidelity for initial states that are superpositions of components with quantum Hamming weights $s_1$ and $s_2$. It can be noticed that the efficiency of quantum state protection is dominated by the maximal Hamming weight, with the minimal fidelity occurring for input states having the highest Hamming weight $s=N$.
If the initial state $\ket{\psi}_{\text{ini},s_1 \oplus s_2}$ 
has a Hamming weight lying between $s_1$ and $s_2$, (with $ s_1 < s_2$), e.g., it is a superposition of states from the two subspaces $\mathcal{H}_{s_1,N} $ and $\mathcal{H}_{s_2,H}$, then the fidelity satisfies
\begin{equation}
    F_{s_2,N}<\bar{F}_{s_1\oplus s_2,N}<F_{s_1,N},
\end{equation}
where $F_{s,N}$ denotes the fidelity for an $N$-qubit input state with quantum Hamming weight $s$, and $\bar{F}_{s_1\oplus s_2,N}$ is the average fidelity for the initial state $\ket{\psi}_{\text{ini},s_1 \oplus s_2}$.  Consequently, the worst-case fidelity under AD noise with single-control filtration is  $F_{N,N}$, which thus constitutes the lower bound of quantum state protection.

\section{Filtration Circuit with Multiple Control Qubits}
\begin{figure}[h]
    \centering
    \includegraphics[width=0.6\columnwidth]{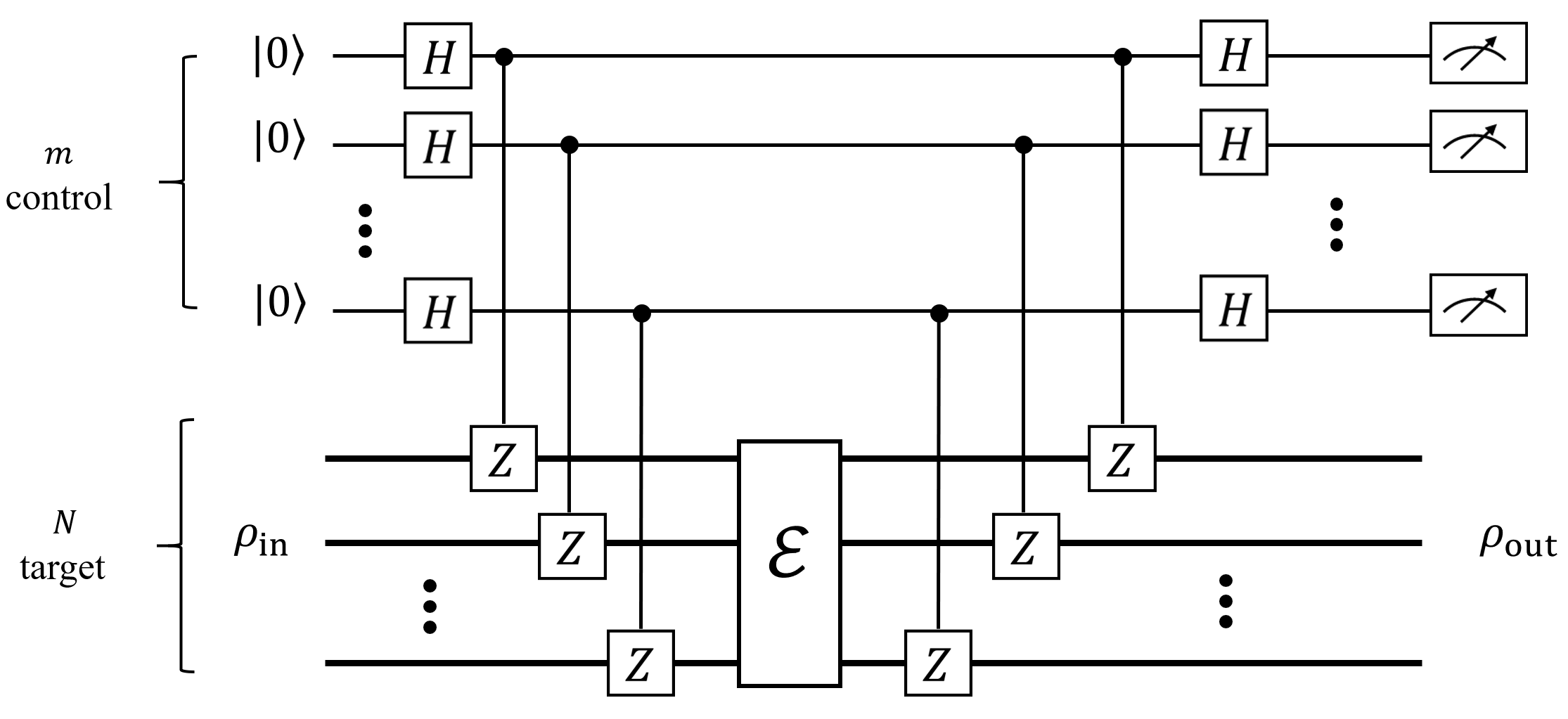}
    \caption{A sketch of the filtration circuit with $m$ control qubits.}
    \label{fig:multi_ctr}
\end{figure}

Although quantum states with the maximal quantum Hamming weight receive the least protection against AD noise under the single-control filtration circuit, the global permutation symmetry raises the question of whether further protection of these states against AD noise can be achieved by introducing additional control qubits. The filtration circuit with multi-control is sketched in Fig.~\ref{fig:multi_ctr}. 

\begin{figure}[h]
    \centering
    \includegraphics[width=0.8\columnwidth]{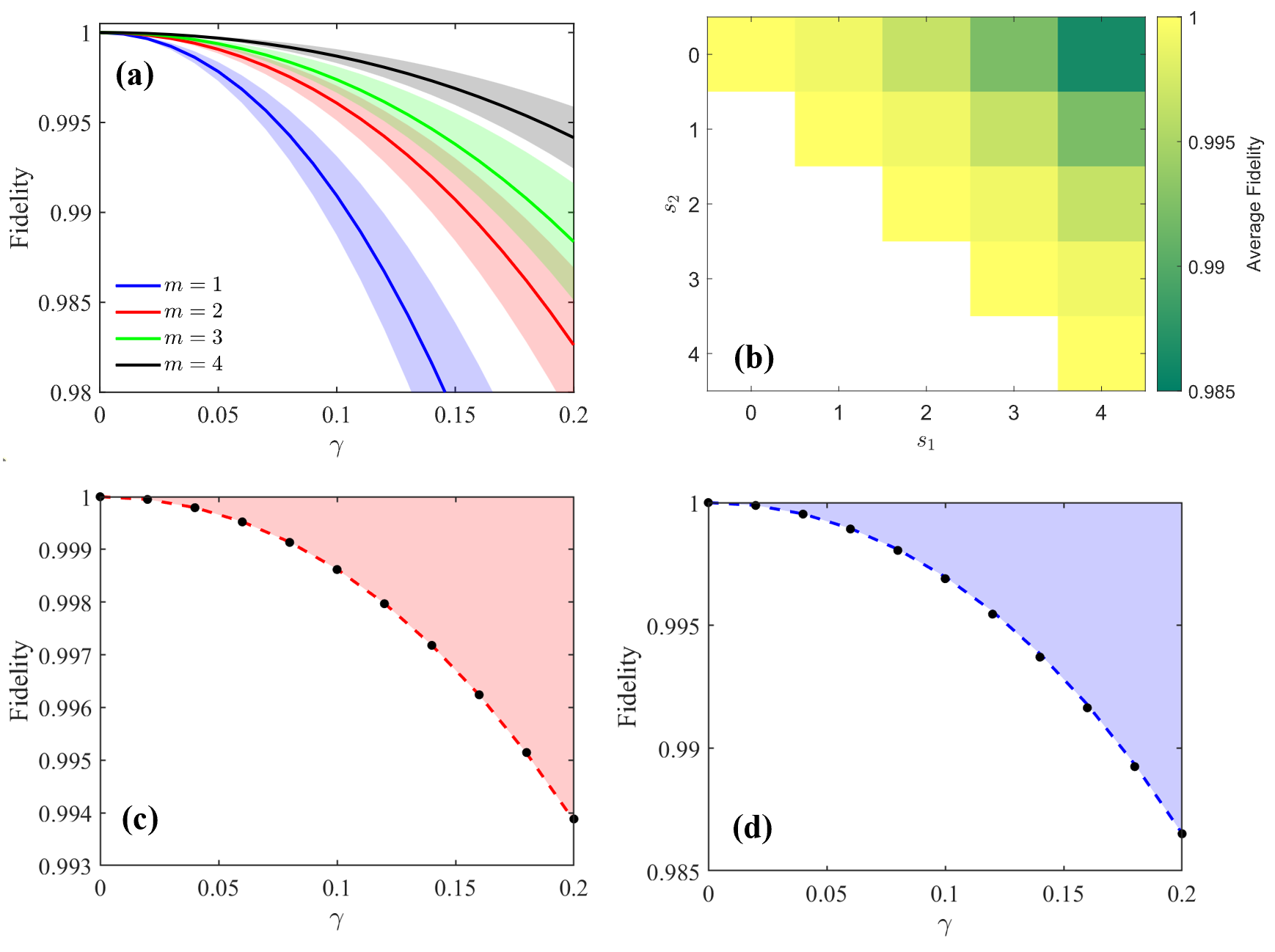}
    \caption{(a) Average fidelity for the multi-control filtration circuit; the input states are sampled from the entire Hilbert space with $N=4$ qubits. The solid lines show the average fidelity for different numbers of control qubits $m$,
    while the shaded regions reflect the standard deviations. (b) Average fidelity of bipartite superposition states with quantum Hamming weights $s_1$ and $s_2$. The diagonal terms correspond to $F=1$, 
    which agrees with the results discussed in the main text. 
    (c) and (d) Fidelity for the 2- and 3-qubit states, respectively.
 In both figures, the shaded regions correspond to the initial states sampled from the respective state spaces; 
 the dashed lines represent the fidelity evaluated for the Bell state and the GHZ state, and the black dots mark the minimum values obtained from the sampling.}
    \label{fig:Nctr}
\end{figure}

Before further discussion, we first define a composite operation that consists of multiple CZ gates applied to different subsets of qubits
\begin{equation}
    \widetilde{CZ}= \bigotimes_{j=1}^m CZ_j,
\end{equation}
where $m$ is the total number of control qubits, 
and each $CZ_j$ represents the $j$-th multi-target CZ gate,
in which the $j$-th control qubit controls a corresponding set of target qubits, given by
\begin{equation}
    CZ_j = \ket{0}\bra{0}_j \otimes \mathbf{I} + \ket{1}\bra{1}_j \otimes \tilde{Z}_j.
\end{equation}
Here $\tilde{Z}_j$ denotes the target Pauli-Z operator controlled by the $j$-th control qubit, which may act on either a single target qubit or multiple target qubits. $\mathbf{I}$ indicates the identity operator.
Without loss of generality, we define
$\bigotimes_{j=1}^m \tilde{Z}_j = \tilde{Z}$, where $\tilde{Z}$ denotes the overall target Pauli-Z operator acting on all data qubits. 
The numerical results of state protection using the multi‑control filtration circuit are plotted in Fig.~\ref{fig:Nctr}(a), where the average fidelity is evaluated for input states uniformly sampled from the entire Hilbert space of $N=4$ qubits. The solid lines represent the average fidelity for different numbers of control qubits $m$, while the shaded regions reflect the corresponding standard deviations. It can be observed that the average fidelity for arbitrary initial states can be consistently improved by increasing the number of control qubits.

As the most favorable case for state protection in terms of fidelity improvement, we consider here the setup with $N$ control qubits, each controlling one of the $N$ target qubits.

In this scenario, the filtration operations acting on all data qubits are identical. Consequently, the circuit involving \(N\) control qubits for \(N\) target qubits can be decomposed into a tensor product of \(N\) independent single-control–single-target filtration processes, and the output state therefore takes the form  
\begin{equation}
    \rho_{\text{out}} = \frac{1}{4^N p_{\vec{0}}} \bigotimes_{i=1}^N \left( \mathcal{E}^{\text{AD}}_i[\rho_{\text{in},i}] + Z_i \mathcal{E}^{\text{AD}}_i[Z_i \rho_{\text{in},i}] + \mathcal{E}^{\text{AD}}_i[\rho_{\text{in},i}Z_i]Z_i + Z_i \mathcal{E}^{\text{AD}}_i[Z_i \rho_{\text{in},i} Z_i]Z_i \right), \label{eq:output_Ncontrol}
\end{equation}
where $p_{\vec{0}}$ is the success probability that all ancillas are measured in the state $\ket{0}$.

We proceed by classifying the input states according to their quantum Hamming weight. Due to the fidelity invariance established in Eq.~\eqref{eq:fid_invariant} and the permutation symmetry of the $N$-control-qubit filtration circuit, all input states sharing the same Hamming weight exhibit identical filtration fidelity under AD noise. Consequently, the protection of an arbitrary state in this scenario can be analyzed by considering a simple representative initial state. A convenient choice for the N-qubit states with Hamming weight $s$ is the product state $\ket{\phi} = \ket{1}^{\otimes s} \otimes \ket{0}^{\otimes (N-s)}$, where the first $s$ qubits are set to $\ket{1}$ and the remaining $N-s$ qubits are set to $\ket{0}$. Substituting this input state into Eq.~\eqref{eq:output_Ncontrol} as
\begin{equation}    
\rho_{\text{in},i}=
    \begin{cases}
    \ket{1}_i\bra{1}, & 1\leq i\leq s, \\
    \ket{0}_i \bra{0}, & s < i \leq N,
    \end{cases} \label{eq:localrho_Ncontrol}
\end{equation}
one can obtain the post-selected state as
\begin{equation}
    \rho_{\text{out}} = \frac{1}{p_{\vec{0}}} (1-\gamma)^s \ket{\phi}\bra{\phi},
\end{equation}
which shows that any initial state $\ket{\phi} \in \mathcal{H}_{s,N}$ can be perfectly protected by the $N$-control-qubit filtration circuit, with the corresponding success probability $p_{\vec{0}}=(1-\gamma)^s$. Since all states with a fixed Hamming weight achieve the same fidelity in our scheme, we conclude that all states in the subspace $ \mathcal{H}_{s,N}$ can be perfectly protected against AD noise by introducing a total of $N$ control qubits.

Since the \(N\)-control-qubit circuit completely eliminates the AD noise for any initial pure state with a fixed Hamming weight, 
we infer that the dominant factor preventing perfect protection for superposition states of distinct Hamming weights is the Hamming distance between their components. 
It therefore follows that our filtration circuit with \(N\) control qubits attains the lowest fidelity for the \(N\)-qubit GHZ state.
The numerically calculated filtration fidelities for superpositions of states with Hamming weights \(s_1\) and \(s_2\) are shown in Fig.~\ref{fig:Nctr}(b). 
These results confirm this conclusion, and a detailed proof is given in Appendix~\ref{app:2}.

We now analytically examine the worst‑case scenario, which occurs when the input state is the generalized GHZ state, for which the Hamming distance between the superposed components is maximal. By substituting its density matrix into Eq.~\eqref{eq:output_Ncontrol}, one obtains the post‑selected output state
\begin{equation}
    \rho_{\text{out,GHZ}} = \frac{1}{2p_{\vec{0},\text{GHZ}}} \left\{ \bigotimes_{i=1}^N\ket{0}_i\bra{0} +(1{-}\gamma)^{N/2}(\bigotimes_{i=1}^N \ket{0}_i\bra{1}+\bigotimes_{i=1}^N \ket{1}_i\bra{0} )+(1{-}\gamma)^N \bigotimes_{i=1}^N \ket{1}_i \bra{1} \right\},
\end{equation}
\begin{equation}
    \rho_{\text{out,GHZ}} = \frac{1}{2p_{\vec{0},\text{GHZ}}} 
    [(\ket{0}^{\otimes N}  +(1{-}\gamma)^{N/2} \ket{1}^{\otimes N})
    (\bra{0}^{\otimes N}  +(1{-}\gamma)^{N/2}( \bra{1}^{\otimes N})
    ]
\end{equation}
with the corresponding success probability 
\begin{equation}
    p_{\vec{0},\text{GHZ}} = \frac{1}{2}\left[ 1 + (1-\gamma)^N \right],
\end{equation}
and fidelity 
\begin{equation}
    F_{\text{GHZ},N} = \frac{1+(1-\gamma)^{N/2}}{2 \sqrt{p_{\vec{0},\text{GHZ}}}}.\label{eq:fid_GHZ_N}
\end{equation}

\section{Conlusion and Outlook}
In this work, we introduce and systematically analyze a quantum filter circuit architecture designed to protect quantum states against AD noise. Both single-control and multi-control filtration circuits are examined. Our results demonstrate a clear advantage in state protection fidelity. For a single-qubit state, the single-ancilla filtration scheme not only achieves higher fidelity than the conventional CSWAP-based approach but also requires substantially fewer quantum resources. Extending to many-body quantum systems, we reveal the remarkable property that states possessing a fixed quantum Hamming weight can be perfectly protected against AD noise using only a single ancilla. Furthermore, we established that the worst-case fidelity in our scheme is fundamentally determined by the Hamming distance between superposed states. However, the fidelity for multi-qubit states can be improved further by introducing additional ancillas. These results indicate that quantum filter circuits provide a resource-efficient and versatile means of suppressing AD noise, and they hold practical value for scalable quantum information protection.

\begin{appendix}

\section{Proof of Fidelity invariance (Eq.~\eqref{eq:fid_invariant})}\label{app:1}
As we have introduced in the main text, 
for $\{ \ket{\phi^{(i)}_{s,N}} \}$ being an orthonormal basis of $\mathcal{H}_{s,N}$, 
any input pure state $\ket{\psi} \in \mathcal{H}_{s,N}$ can be expressed as
$\ket{\psi}=\sum_i \alpha_i \ket{\phi^{(i)}_{s,N}}$.
Then for $\psi_{\text{in}} \in \mathcal{H}_{s,N}$, 
the fidelity under AD taks the form
\begin{equation}
    F(\mathcal{E}^{\text{AD}}(\ket{\psi } \bra{\psi}),\ket{\psi } \bra{\psi}) = 
      \bra{\psi} 
    \sum_{m}^{m\leq N} \sum_{k=1}^{\binom{N}{m}} E_{P_{k}(N,m)}
   \ket{ \psi}.
\end{equation}
For $m\neq 0$, $E_{P {(N,m)}}\ket{\psi}\in\mathcal{H}_{s-m,N}$. 
thus, $\bra{\psi}E_{P {(N,m>0)}}\ket{\psi}=0$.
Then one can obtian 
\begin{equation}
   \bra{\psi }E_{P {(N,m)}}\ket{\psi } =
\begin{cases}
\sum_{i,j} \alpha_j \alpha_i^* \bra{\phi^{(i)}_{s,N}}
   E_{P {(N,m)}}
   \ket{\phi^{(j)}_{s,N}}=\sum_i \alpha_i^2
     (1-\gamma)^{s/2} 
 =(1-\gamma)^{s/2},   & m = 0 \\
0,    & m \neq 0.
\end{cases}
\end{equation}
Then for any state  $\rho=\sum_i \lambda_i \ket{\psi_i}\bra{\psi_i}$,
where we choose a suitable orthonormal basis \(\{\ket{\psi_i}\}\) that diagonalizes \(\rho\), 
we have 
\begin{equation}
\begin{split}
    F(\mathcal{E}^{\text{AD}}(\rho),\rho) &=
 F( \mathcal{E}^{\text{AD}}(\sum_i \lambda_i
 \ket{\psi_i } \bra{\psi_i}),\sum_i \lambda_i \ket{\psi_i } \bra{\psi_i}) \\
 &=\sum_i \lambda_i (1-\gamma)^{s/2} 
  =(1-\gamma)^{s/2}.
\end{split} 
\end{equation}
Since the filtering process eliminates the anti-symmetric Kraus operators (\(m = 1, 3, \dots\)), the above analysis can be extended to the final post-selected output states. 
In other words, for all input states \(\rho_{\mathrm{in}} \in \mathcal{D}(\mathcal{H}_{s,N})\), the fidelity between the filtered output state and the input state differs from the fidelity under the bare AD noise only by a factor that depends solely on the measurement success probability.
Consequently, this filtered fidelity is identical for every input state within the subspace \(\mathcal{H}_{s,N}\).

\section{Proof of minimal-fidelity states}\label{app:2}
Here, we prove that in the $N$-ancillary scenario, the worst-case fidelity occurs when the Hamming weight distance is largest, e.g., when the input state is a generalised GHZ state. For simplicity, we assume that the initial input state is an $N$-qubit pure state, which can always be decomposed as
\begin{equation}
    \ket{\Psi} = \bigotimes_{i=1}^k \ket{\psi_i}_{n_i},
\end{equation}
where $\ket{\psi_i}_{n_i}$ indicates the $i$-th tensor component comprising $n_i$ qubits, with $\sum n_i =N$. 
For $k=N$, the initial state $\ket{\Psi}$ is a product state of all \(N\) qubits, whereas for $k=1$, $\ket{\Psi}$ is the maximally entangled state, in which all $N$ qubits are globally correlated. 
For $1<k<N$, the components $\ket{\psi_i}_{n_i}$ are combined via tensor products to form \(|\Psi\rangle\), and each \(|\psi_i\rangle_{n_i}\) itself is partially entangled (\(n_i \geq 2\)).
Under this decomposition, the overall filtration can be decomposed into \(k\) independent filtering processes, each using \(n_i\) ancillary qubits as controls and acting on the corresponding \(|\psi_i\rangle_{n_i}\).

Since the fidelity of tensor product states satisfies the multiplicative property
\begin{equation}
    F(\bigotimes_i \rho_{1,i},\bigotimes_i \rho_{2,i}) = \prod_i F(\rho_{1,i},\rho_{2,i})
\end{equation}
it follows that for an \(N\)-qubit product state \(|\psi\rangle\), 
the fidelity after purification is 
\begin{equation}
    F = F_{i}^{N} \geq \left(\frac{1+\sqrt{1-\gamma}}{\sqrt{4-2\gamma}}\right)^{N}\sim 1-\frac{N}{32}\gamma^2 + O(\gamma^3)
    \geq F_{\text{GHZ},N} \sim 1-\frac{N^2}{32}\gamma^2 + O(\gamma^3),
\end{equation}
where \(F_{i}\) denotes the $i$-th single-qubit fidelity given in Eq.~\eqref{eq:fidelity_1qubit}.
For partial entangled states, the fidelity satisfies 
\begin{equation}
\begin{split}
    F &=\prod_{i} F_{i} \prod_j F_{j,\text{ent}}
    \geq [1-\frac{n_i}{32}\gamma^2 + O(\gamma^3)]
    \prod_j [1-\frac{n^2_j}{32}\gamma^2 + O(\gamma^3)] \\
    &\geq \prod_j [1-\frac{n^2_j}{32}\gamma^2 + O(\gamma^3)]
    \sim 1- \frac{1}{32} \sum_j n^2_j \gamma^2 + O(\gamma^3)\\
        & \geq 1 - \frac{1}{32}(\sum_j n_j)^2 \gamma^2 + O(\gamma^3) = F_{\text{GHZ},N}.
\end{split}
\end{equation}
where the index \(i\) labels the product-state components and \(j\) labels the entangled states.
Hence, the proof is complete.

\end{appendix}

\bibliography{reference}

\end{document}